**Stimulated Raman with Broadband LED Stokes Source for Analysis of Glucose**

Peter Bullen [1], Ioannis Kymissis [1], Adler Perotte [2]

[1] *Columbia University, 500 W 120$^{th}$ St, New York, NY 10027*

[2] *Columbia University Medical Center,* 617 West 168th St., New York, NY 10032

psb2125@columbia.edu

We demonstrate stimulated Raman gain using a broadband LED Stokes source to measure vibrational spectra of aqueous glucose solutions. This versatile and cost-effective method increases Raman signal for a variety of applications. We measured both stimulated Raman and spontaneous Raman spectra of glucose solutions with concentrations up to 10 mM with a photon counter and lock-in amplifier. We built partial least squares regression models based on both stimulated Raman and spontaneous Raman spectral data measured with each instrument for predicting concentrations of the glucose solutions. The stimulated Raman spectra measured with the lock-in amplifier based model had the strongest predictive power and predicted the concentrations of the test set of glucose solutions with a mean squared error value an order of magnitude lower than those of the spontaneous Raman based model.

**Introduction**

Stimulated Raman Scattering (SRS) spectroscopy is a powerful technology for high sensitivity quantitative analysis of molecular vibrations. Raman spectra are determined by the fundamental vibrational modes of molecules, and therefore are highly specific compared spectra generated by other spectroscopic techniques such as NIR absorption spectroscopy.[1] Furthermore, SRS provides a significant signal enhancement compared to the relatively weak spontaneous Raman signal, enabling detection of molecules in highly dilute solutions and a variety of other high sensitivity applications.[2] SRS involves two light sources: pump and Stokes. When the frequency difference between the pump and Stokes sources corresponds to one of the Raman vibrational modes of the molecules in the sample, the molecular transition probability greatly increases.[3] This technique results in intensity gain at the

Stokes frequency, Stimulated Raman Gain (SRG), and intensity loss at the pump frequency, Stimulated Raman Loss (SRL).[4] Both the SRG and SRL signals can be orders of magnitude higher than the spontaneous Raman signal, and either may be used for molecular detection or imaging.[5,6]

The increased stimulated Raman signal enables high sensitivity detection and high contrast imaging with fast integration times and low laser excitation powers, all of which are critical to biomedical applications. Due to the combination of these beneficial attributes, SRS microscopy has been used to great effect for imaging biological issue in vivo at video-rate speeds.[7] In recent years, researchers have developed a wide variety of powerful SRS-based biomedical technologies including three-dimensional spectral imaging of proteins,[8] in vivo brain tumor imaging,[9] and a handheld in vivo SRS microscope.[10] SRS also holds several advantages compared to Coherent Anti-Stokes Scattering (CARS) spectroscopy, another coherent Raman technology often used for sensitive high-speed imaging. While CARS can be obscured by non-resonant background and autofluorescence, SRS is unaffected by these phenomena.[11,12,13] Depending on the modulation method used, SRS can reach higher levels of sensitivity, limited only by shot noise.[14] Finally, SRS spectra are identical to spontaneous Raman spectra, which enables easier analysis and comparison to measured spontaneous Raman spectra in the literature.[15]

One limitation of the standard SRS system is that it can only measure the SRG or SRL signal corresponding to a single Raman-active molecular vibration at a time: that with a resonant frequency equal to difference between the pump and Stokes frequencies. This presents a challenge for applications requiring a broader Raman spectrum. Quantitative analysis of complex multi-component samples often relies on a specific vibrational mode, separate from the mode of interest, to serve as a reference measurement or internal standard. For example, in vivo blood glucose level monitoring has been demonstrated via spontaneous Raman spectroscopy with the characteristic glucose mode at 1125 $cm^{-1}$ being the vibrational mode of interest and the hemoglobin mode at 1549 $cm^{-1}$ serving as the internal standard.[16] In order to perform similar types of full spectrum measurements with the benefit of SRS enhanced signal, researchers have developed innovative SRS systems based on tunable pump or Stokes sources which scan across the Raman spectrum,[17,18] tunable optical bandpass filters which filter continuum laser sources down to scanning narrowband pump or Stokes beams,[19-21] and femtosecond broadband stimulated Raman spectroscopy (FSRS).[22-24] FSRS in particular has shown great promise for high speed spectroscopic applications such as monitoring chemical reactions in real time due to its high SNR, temporal resolution, and bandwidth. To further reduce noise generated by jitter between the pump and Stokes beam, an FSRS system in which both the narrowband Stokes and continuum pump originate from the same titanium-sapphire laser oscillator has been demonstrated.[25] One drawback of FSRS and many SRS systems in general is the high cost and lack of portability of major components such as femtosecond lasers. Some spectroscopic applications, such as the non-invasive blood monitoring mentioned previously, would benefit from the broadband signal enhancement of FSRS, but must also be cost-effective and adaptable to suit real-world medical environments. For applications in which only a moderate Raman signal enhancement is required, a cost-effective broadband SRS solution is highly desirable.

In this article, we demonstrate a broadband SRS system which uses a high-power LED as the continuum Stokes source and a cw laser pump. Due to the lower peak power of LEDs compared to pulsed laser sources, this system enhances the Raman signal with lower efficiency than the traditional SRS system involving two laser sources. However, this prototype system allows for modest SRG of all Raman modes over the spectral width of the LED, while reducing cost and complexity. We test the capability and limitations of this SRS system by measuring Raman spectra of aqueous glucose solutions. We chose to measure glucose solutions as our test samples due to the recent interest in developing non-invasive blood glucose monitoring.[16,26,27] In order to take advantage of the full spectrum data and modest enhancement, we use the measured spectra to build partial least squares regression (PLS) models able to predict the glucose concentration of a solution for its Raman spectrum. Even the modest signal enhancement from our SRS system is valuable for building predictive statistical models such as PLS because data across the entire spectrum contributes to the model, leading to more accurate predictions.[28,29] We build PLS models from both spontaneous and stimulated Raman spectral data and compare their predictive power for glucose solution concentration.

**Experimental Setup**

In order to make a fair comparison between broadband stimulated Raman with LED Stokes source and standard spontaneous Raman, we built an optical system capable of employing either spectroscopic technology without any modifications to the optical components. The only difference between the two techniques is that stimulated Raman involves two light sources: both the LED and laser. Regardless of the spectroscopic method, the optical system directs light to the liquid sample, collects and filters the scattered light, and detects the Raman scattered light with a photomultiplier tube (Fig. 1). The photomultiplier tube is connected to one of two instruments for the final signal measurement: either a photon counter or lock-in amplifier, and measurements were taken with both instruments for both spontaneous and stimulated Raman for comparison. Optical fibers are used to connect some parts of the optical setup to each other in order to simplify system optimization as adjustments to one portion of the system do not cause cause misalignment in other parts connected via optical fibers.

For both spontaneous and stimulated Raman spectroscopy, the 532 nm pump (excitation) laser is split by a 10:90 beam splitter, and the smaller portion is reflected towards a photodiode and used as an intensity reference to compensate for laser drift. The transmitted portion of the beam is directed through an aperture stop and laser line filter to clean up the spatial profile of the beam and attenuate any spectral energy apart from the 532 nm laser line. The laser power is 200 mW at the source and is about 150 mW upon reaching the sample. A longpass dichroic beam splitter, oriented 45° with respect to the incident beam, is used to reflect the pump beam to the sample while permitting the LED Stokes beam to be transmitted through towards the same point in the sample. This is possible because the dichroic reflects over 94% of incident light at or below 532 nm, including the pump beam, and transmits over 93% of light above 541.6 nm, which corresponds to most of the Stokes Beam so that it may stimulate Raman modes with frequency shift of 333 cm$^{-1}$ or greater. The pump and Stokes beams are focused into the volume of the liquid sample by an aspheric lens. Scattered light is collected in the forward direction by the a low f-number aspheric lens, and the low f-number of 1 and high numerical aperture of 0.5 ensures high collection efficiency. The collected light is collimated and coupled into an optical fiber via coupling lens, further filtered by a dielectric longpass filter, and directed into the monochromator. The longpass filter and monochromator both attenuate the intense Rayleigh scattered light from the pump beam to prevent stray light from affecting the final spectra as much as possible. The entrance and exit slits of the monochromator are set at 100 microns to insure enough light passes through while maintaining sufficient spectral resolution. The monochromator must be adjusted manually to measure the Raman signal at each wavelength.

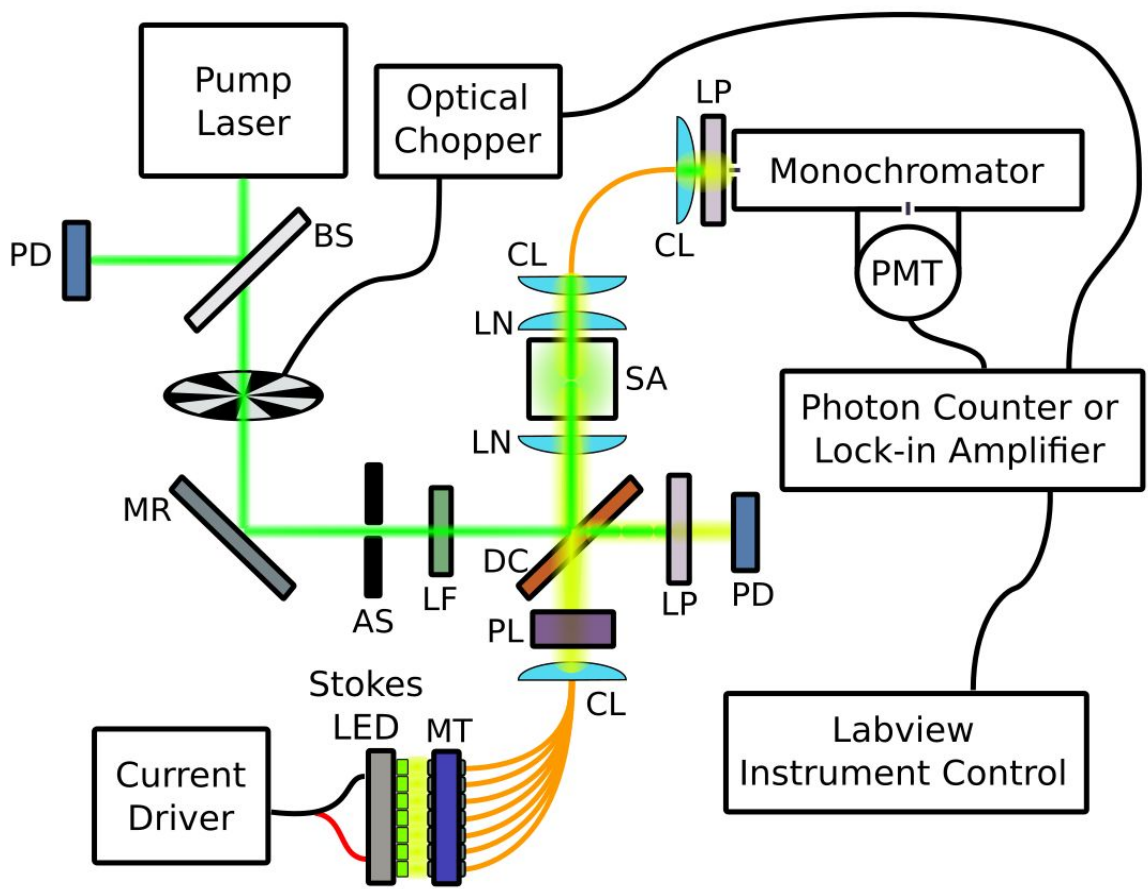

Figure 1: Stimulated Raman Spectroscopy system diagram. The beams from the excitation laser and LED are combined via dichroic beam splitter and focused into the sample. Scattered light is ultimately detected by Photomultiplier Tube (PMT) and measured using either a photon counter or lock-in amplifier. The optical chopper is only used in conjunction with the lock-in amplifier and provides the reference frequency. PD = photodiode, BS = beam splitter, MR = mirror, AS = aperture stop, LF = line filter, DC = dichroic beam splitter, LN = lens, SA = sample, CL = collimating/coupling lens, LP = longpass filter, PL = polarizer, MT = optical fiber mount

Light passing through the monochromator is detected by the photomultiplier tube (Hamamatsu R4220P) and the signal is measured by either the photon counter (Stanford Research Systems SR400) or Lock-in Amplifier (Stanford Research Systems SR810). Our photomultiplier tube is optimized for visible light measurements, which is one of the reasons we chose to use visible pump and Stokes sources. The photomultiplier tube is reverse biased at 1.1 kV, a value chosen to optimize sensitivity without introducing excess noise. The photomultiplier tube is wrapped in layers of aluminum foil to block out stray light from the room. Measurements are taken with the room lights out to further minimize stray light, and the room is cooled to about 15 °C, resulting in a low dark current of 0.48 counts per second measured by the photon counter. Both the photon counter and lock-in amplifier data collection is automated with Labview software.

For stimulated Raman measurements, the Stokes beam is generated by seven high powered LEDs centered at 567 nm. The LEDs are wired together in series and soldered to a single aluminum PCB base with a heat sink for temperature stabilization. The optical power from the LEDs is about 550 mW and is collected by seven ends of a multi-furcated optical fiber connected to a 3D-printed optical fiber mount. The high power is necessary for efficient SRS, especially due to the significant loss inherent to coupling incoherent light into optical fibers. The Stokes beam enters the main portion of the optical system via the multi-furcated optical fiber from the direct opposite side of the dichroic from the sample. The beam is collimated and vertically polarized to match the polarization of the pump beam for optimal stimulated Raman gain. The collimated and polarized beam is transmitted through the dichroic and focused into the sample through the same lens as the pump beam in order to increase spatial overlap between the focal volumes of the two light sources. At the position of the sample, the Stokes beam power is about 10 mW. A small portion of the beam is reflected by the dichroic beam splitter towards a photodiode and used as an intensity reference. A small portion of the pump beam is also transmitted through the dichroic and is filtered out by a longpass filter so the photodiode monitors only the Stokes beam. By monitoring a reference intensity for both pump and Stokes sources, we found that after turning the light sources on, the intensity decreased over the course of several hours by about 5% and 15% respectively before stabilizing. To further mitigate the effect of intensity drift, we took all measurements a minimum of 24 hours after turning on both light sources.

The aqueous glucose solutions are contained in transparent cuvettes (Eppendorf UVette). The path length through the cuvette is 1 cm, and volume of the solutions is 2 mL. The glucose solutions are composed of DI water and pharmaceutical grade D-glucose (Sigma Aldrich). The cuvettes are mounted in a 3D-printed mount to insure precise and consistent positioning between the two lenses to minimize any systematic error between different samples due to cuvette position.

**Stimulated Raman Measurements**

When taking stimulated Raman measurements with the photon counter, two measurements are required for each data point: a combined signal ($I_{com}$) in which both the pump and Stokes beams are present and the Stokes signal ($I_{Stokes}$) in which the pump is blocked. The Stokes measurements were taken immediately after the combined measurement by blocking the pump laser with a manual shutter. The Raman gain, defined as the quotient of the combined and Stokes signals, is exponentially proportional to the concentration of the Raman-active molecule and the pump signal,[22] as shown in equation (1),

$$Raman\ gain = \frac{I_{com}}{I_{Stokes}} = \exp(a\ \sigma_R c\ z\ I_{pump}) \tag{1}$$

where $a$ is a proportionality constant, $\sigma_R$ is the Raman cross section, $c$ is the concentration, and $z$ is the focal depth in which the beams spatially overlap. We tested our stimulated Raman system on methanol solutions for this relationship between pump intensity and Raman gain. We choose methanol for our test solution because it has greater Raman response than the glucose solutions and has a prominent Raman mode at 565 nm (1062 cm$^{-1}$), which is close to one of the main glucose Raman modes at 567 nm (1120 cm$^{-1}$). We increased the pump power from 200 mW to 1 W and found that the Raman gain increased exponentially with the pump power (Fig. 2).

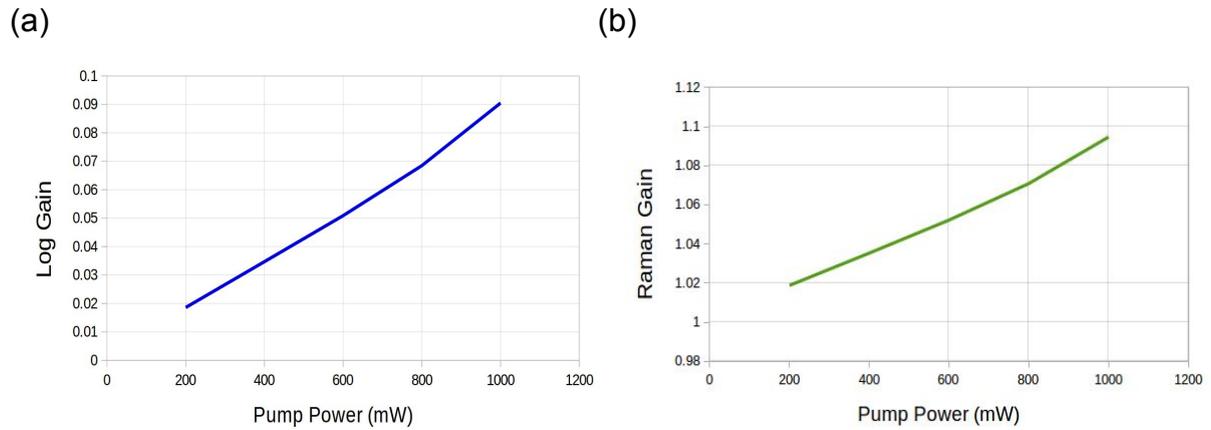

Figure 2: (a) Natural logarithm of Raman gain as a function of pump power for 1062 cm$^{-1}$ mode in methanol. The natural logarithm follows a linear trend with respect to pump power with R$^2$ value of 0.995, indicating that the variation between Raman gain and pump power may be modeled by exponential fit. (b) However, the Raman gain itself also follows a linear trend with respect to pump power with R$^2$ value of .993, indicating that the Raman gain may appropriately be modeled by its first-order expansion in equation (3) at these pump power levels.

However, in most stimulated Raman experiments, including ours, which use low to moderate power levels for the pump beam, the gain is relatively small, and equation (2) is satisfied

$$y = a\,\sigma_R\,c\,z\,I_{pump} \ll 1 \tag{2}$$

so we can approximate the exponential as its first-order expansion (3).

$$e^y \approx 1 + y \tag{3}$$

To make the significance of our results more apparent, we define the difference between the combined and Stokes measurements to be the stimulated Raman gain signal ($\Delta I_{SRG}$) (4).

$$\Delta I_{SRG} = I_{com} - I_{Stokes} \tag{4}$$

Then, substituting equations (3) and (4) into equation (1), we arrive at equation (5):

$$\frac{I_{com}}{I_{Stokes}} = \frac{I_{Stokes} + \Delta I_{SRG}}{I_{Stokes}} \approx 1 + a\,\sigma_R\,c\,z\,I_{pump} \tag{5}$$

which simplifies to

$$\Delta I_{SRG} \approx a\,\sigma_R\,c\,z\,I_{pump}\,I_{Stokes} \tag{6}$$

showing that the SRG signal is proportional to both the pump and Stokes intensities in our small signal approximation.

All stimulated Raman spectra measured using the photon counter were calculated as a difference between the combined and Stokes signals for a particular glucose solution sample (Fig. 3) according to equation (4) instead of as the quotient in equation (1) in this report. This approximation is made in order to make comparing spectra more straightforward as both spontaneous and SRG difference spectra share the same units of counts while the Raman Gain quotient is a unitless quantity. Furthermore, the measurements taken by the lock-in amplifier are essentially difference measurements as the lock-in signal is proportional to the amplitude of the component of the input signal varying at the frequency of the optical chopper. Although the SRG difference signal is proportional to the Stokes signal which varies with frequency shift, this did not significantly affect the predictive power of our partial least squares regression models.

(a)
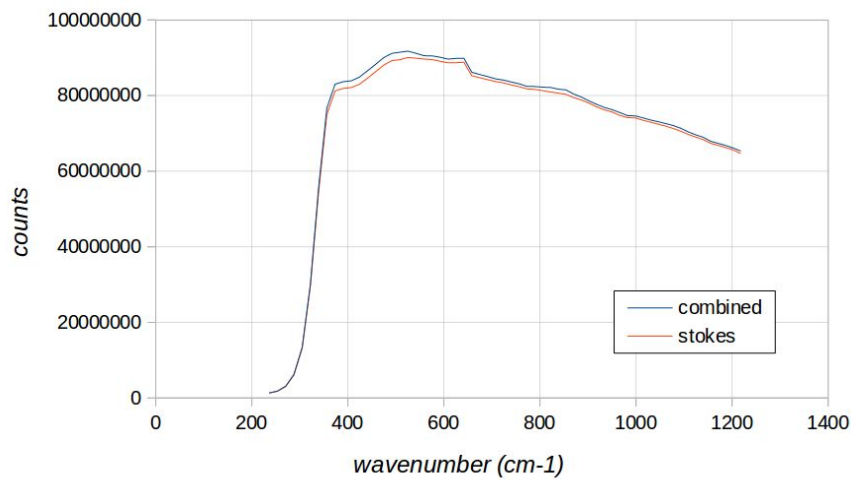

(b)
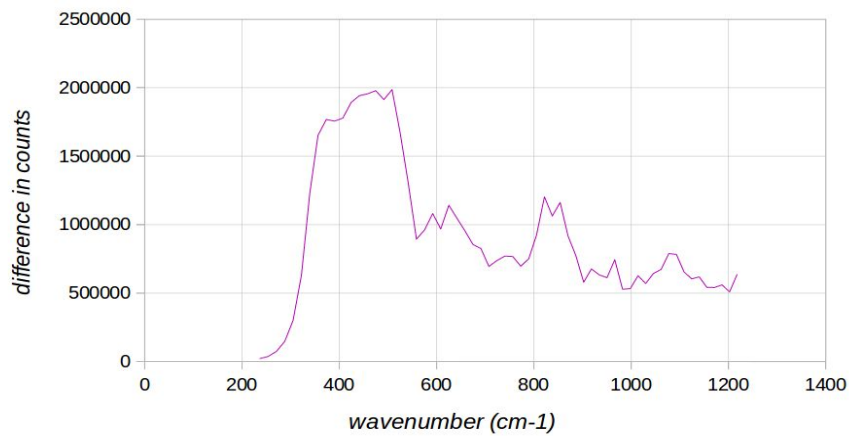

Figure 3: (a) Combined (pump laser and Stokes LED) and Stokes (only Stokes LED) signals for 10 mM glucose sample measured with the photon counter. (b) Raw

Stimulated Raman Gain signal is approximated as the difference between the combined and Stokes signals.

In order to build robust numerical models that account for intensity drift of the pump laser and Stokes LED throughout the data collection process, intensity references were taken for both light sources with photodiodes for each data point. Since the stimulated Raman gain is proportional to both the pump and Stokes intensity, each stimulated Raman gain measurement was divided by both the pump and Stokes reference taken for that point before the spectra were used by the partial least squares model. In order to compare these processed normalized Raman spectra to the spontaneous Raman spectra, each spontaneous measurement was divided by the pump reference for that point and the average of the Stokes references for the stimulated measurements since the spontaneous measurements do not involve the Stokes LED but must still be divided by an average Stokes reference for fair comparison to the normalized stimulated spectra. This normalization process improved the predictive power of our partial least squares models overall with the normalization to the pump reference being more significant than that of the Stokes reference.

**Experimental Process**

We measured the spontaneous and stimulated Raman spectra of aqueous glucose solutions with our optical setup. The solutions were contained in cuvettes and glucose concentrations ranged from 10 mM to 0 mM (only DI water). This range is useful for study because normal human blood glucose concentration averages at about 5.5 mM. Using both the photon counter and lock-in amplifier, we collected both spontaneous and stimulated Raman spectra for three separate glucose solutions at each concentration level. This would allow us to conveniently divide the data into training, validation, and test sets later when analyzing the data. We collected data from 540 to 570 nm on our monochromator for each sample, which resulted in Raman spectra from 289 to 1262 cm$^{-1}$ (conversion to wavenumbers includes an instrument-specific calibration offset). Measurements were taken one data point at a time every 0.5 nm from 540 to 570 nm, which accounts for most of the significant spectral features of glucose. The integration time for each measurement was 5 seconds using the photon counter and the time constant for the lock-in amplifier was 3 seconds. The average dark current of 2.4 counts per 5 seconds was subtracted from each photon counter measurement.

**Results**

The glucose solution spectra measured via spontaneous and stimulated Raman spectroscopy methods show similar spectral features corresponding to aqueous glucose Raman modes investigated in the literature.[30,31] Due to using a monochromator and taking one data point rather than a full spectrum at a time, our spectra have relatively low resolution, but be can identify groups of Raman modes that likely correspond to a particular spectral feature. The spectral feature around 1092-1139 $cm^{-1}$ likely corresponds to the COH bending mode characteristic of glucose, that between 853-918 $cm^{-1}$ is the combination of C-H and C-C bending and stretching modes of both alpha and beta glucose anomers, that around 675-724 $cm^{-1}$ is composed of deformations in the ring, and the large group of features between 424-575 $cm^{-1}$ is likely the combination of C-C-O deformation modes and endocyclic vibrational modes of the ring. The normalized Raman intensity is generally higher for the stimulated Raman measurements compared to the spontaneous Raman measurements throughout the spectra for both measurements taken with the photon counter and with the lock-in amplifier (Fig. 4). This suggests that the the pump laser is interacting with the Stokes LED and causing stimulated Raman gain in the glucose solutions. It is likely that our stimulated data are result of both the spontaneous and stimulated Raman effects scattering light simultaneously, but here we are concerned only with the total signal measured using both the pump and Stokes LED and not what portion of this signal is generated by the stimulated Raman effect.

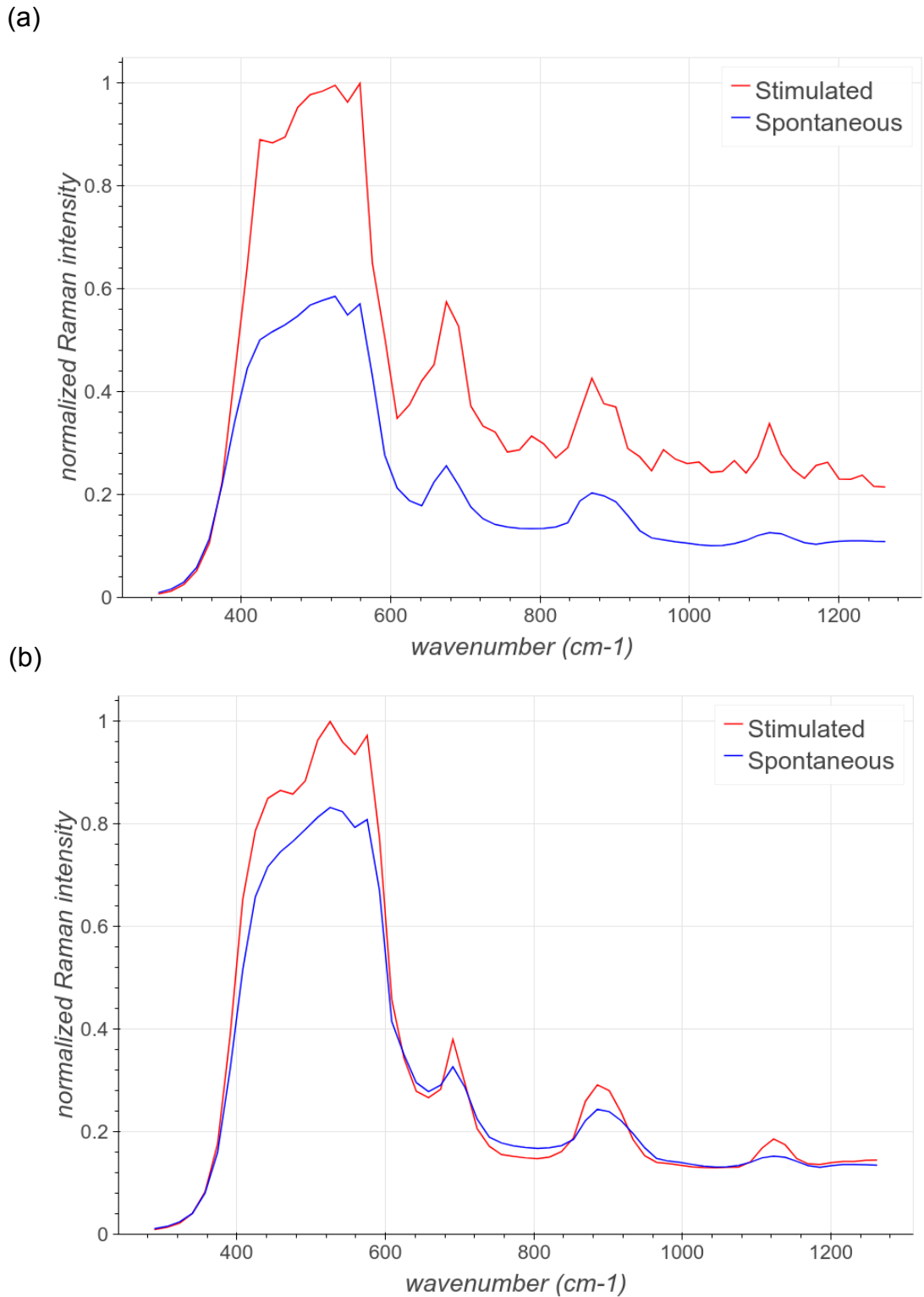

Figure 4: Comparison of stimulated and spontaneous Raman spectra of a 10 mM glucose solution measured with (a) photon counter and (b) lock-in amplifier. The enhancement between stimulated and spontaneous spectra is greater with the photon counter spectra, but the stimulated photon counter spectra also has significantly higher noise.

The relative intensity increase between stimulated and spontaneous Raman was greater for for the measurements taken with the photon counter than those taken with the lock-in amplifier. For the photon counter, the stimulated measurements were 1.5-3 times higher than the spontaneous measurements across the spectra for a given glucose solution. However, this enhancement factor was not strongly correlated with the Raman modes, so the relative increase between stimulated and spontaneous measurements was similar for both Raman-active and non-Raman-active regions of the spectra. In contract, the stimulated measurements taken with the lock-in amplifier were 1.05-1.3 times only higher than the spontaneous measurements, but this enhancement factor was strongly correlated with the Raman modes. The enhancement factor was 1.2-1.3 around the Raman-active regions while only 1.05-1.15 for the non-Raman-active regions. The relative difference between the intensity of Raman peaks and the non-Raman-active background is generally more critical than absolute intensity for numerical analysis, so the stimulated lock-in measurements are at an advantage here.

Another weakness of our photon counter stimulated Raman measurements is noise. Across all spectra, the average coefficient of variation, a measure of relative noise and defined as the quotient of the standard deviation and mean, is 16% for photon counter stimulated, 0.38% for photon counter spontaneous, 0.29% for lock-in stimulated, and 0.31% for lock-in spontaneous measurements. The photon counter stimulated Raman measurement noise is almost 2 orders of magnitude higher than all other measurement techniques, which can be confirmed qualitatively by the noiser photon counter stimulated Raman spectra (Fig. 5). While the lock-in amplifier did reduce the noise slightly from 0.38% to 0.31% for the spontaneous measurements, it is clearly most useful in reducing the stimulated measurement noise from 16% to 0.29%. While monitoring the reference intensities of both the pump laser and Stokes LED compensated for intensity drift to some extent, the Stokes signal alone measured by the photon counter was almost 2 orders of magnitude across the spectrum, so any fluctuations in the LED would have a relatively large effect on the stimulated Raman signal. Overall, the additional noise introduced by the LED is a greater detriment to the stimulated photon counter data than any signal enhancement for numerical analysis.

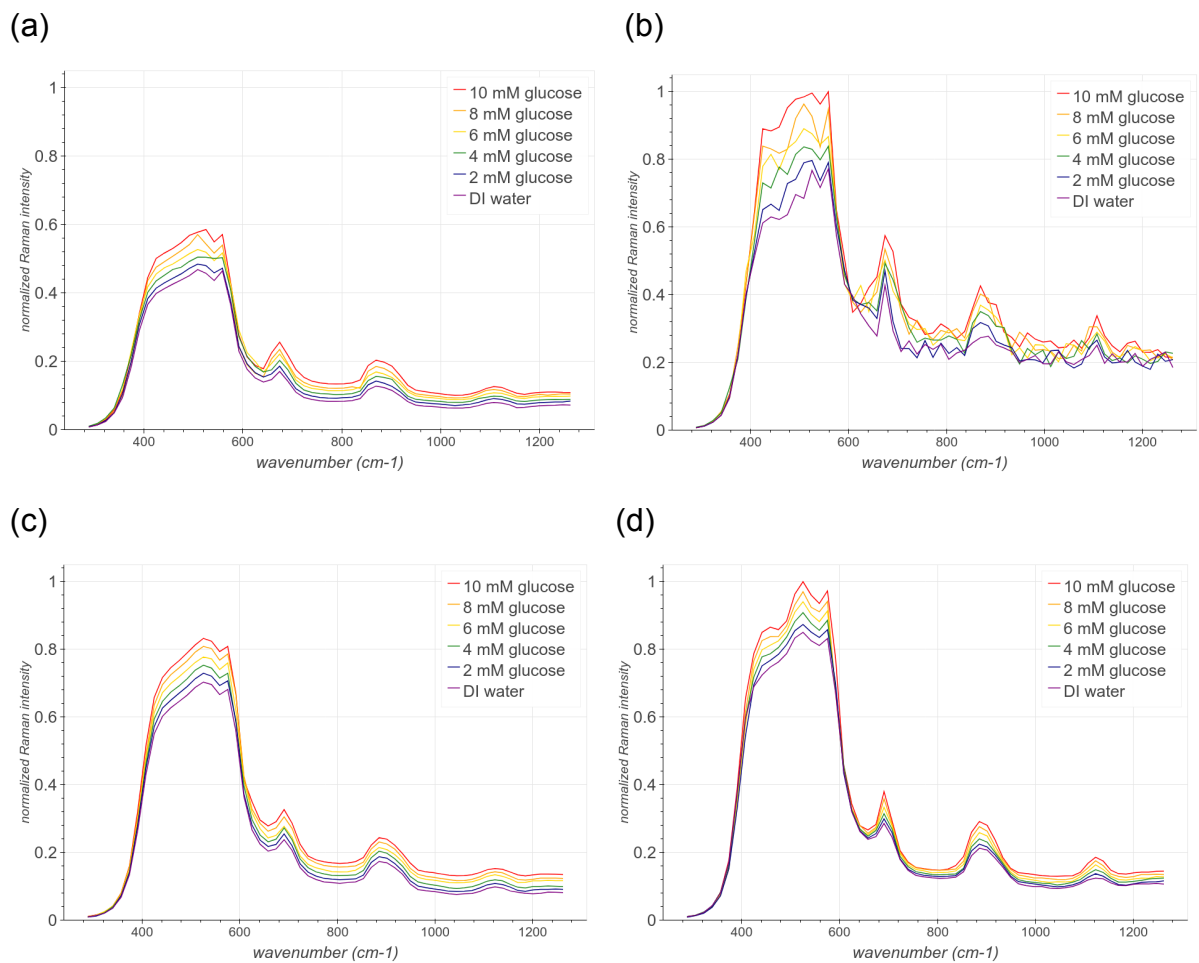

Figure 5: (a) Photon counter spontaneous, (b) photon counter stimulated, (c) lock-in spontaneous, (d) lock-in stimulated Raman spectra of glucose solutions with concentrations from 0-10 mM. Both sets of photon counter spectra are normalized to the highest point of the photon counter stimulated data, and both sets of lock-in spectra are similarly normalized to the highest point of the lock-in stimulated data. Differentiation between the glucose solutions of different concentration can be seen for all measurement techniques. The relative Raman intensity is generally higher for the stimulated spectra compared to the respective spontaneous spectra, but the enhancement factor between stimulated and spontaneous measurements is higher for the photon counter measurements overall. The photon counter stimulated spectra are significantly noisier than the other spectra. The stimulated lock-in spectra show slightly narrower spectral features than the corresponding spontaneous spectra, and Raman modes can be resolved to a slightly higher degree in the 424-575 cm$^{-1}$ region.

Both spontaneous and stimulated Raman spectra of glucose solutions measured with either the photon counter or lock-in amplifier show a significant difference in Raman intensity between the varying glucose solutions (0, 2, 4, 6, 8,10 mM), especially at the group of Raman features between 424-575 cm$^{-1}$ (Fig. 5). This enables any of the four sets of data to be used as a training set for a statistical model, such as partial least squares, which would predict the glucose concentration of an unknown solution given its Raman spectrum. However, the stimulated Raman spectra measured with the lock-in amplifier is the best set of data for building the predictive model because of its enhanced signal and low noise. Compared to the spontaneous lock-in spectra, the stimulated lock-in spectra are slightly enhanced, with the regions immediately surrounding Raman modes showing the greatest signal increase. The enhancement specifically at the glucose Raman modes creates greater differentiation between spectra of different glucose solutions at the regions most strongly correlated with glucose concentration. Although the stimulated photon counter spectra show even greater overall enhancement than the stimulated lock-in spectra, the enhancement factor is less correlated with Raman modes, and more importantly, these spectra have significantly greater noise, making them a weaker data set for building a predictive model.

**Partial Least Squares Analysis**

The partial least squares (PLS) regression is a simple statistical model which can be used for predicting concentrations based on spectral data. In this study, we use PLS to test and compare the predictive power of our spontaneous and stimulated Raman spectral data for predicting concentrations of glucose solutions. We programmed our PLS model in Python using the scikit-learn package [32]. We analyzed our Raman data over the full collected spectral range of 289-1262 $cm^{-1}$, and two smaller portions of the spectrum: 458-625 cm-1 and 642-1129 cm-1. Since we measured three separate glucose solutions for each concentration value, the first set of measurements for all concentrations 0-10 mM was the training set, the second was the validation set, and the third was the test set. In order to find the optimal number of PLS components for the model, we trained and validated our the model using a progressively higher number of PLS components until the mean squared error of the prediction increased, signifying overfitting. Therefore, the previous number of components was optimal. The optimal number of components for the PLS model based on photon counter stimulated spectra was 2 for each spectral range, likely due the greater noise of these spectra. The number of components for the PLS model based on photon counter spontaneous spectra was 3 or 4 depending on spectral range, and that for the lock-in spontaneous and stimulated spectra was 4 or 5 depending on spectral range. Generally, the greater number of components the PLS model can use without overfitting, the stronger the predictive power of the model. Then, the trained and validated model predicted the glucose concentrations based on the test data set, and these predictions were compared to the actual glucose concentration values (Fig. 6). The mean squared error between the predicted and actual glucose concentration measured the predictive power of each model (Table 1), with lower error signifying higher predictive power of the model.

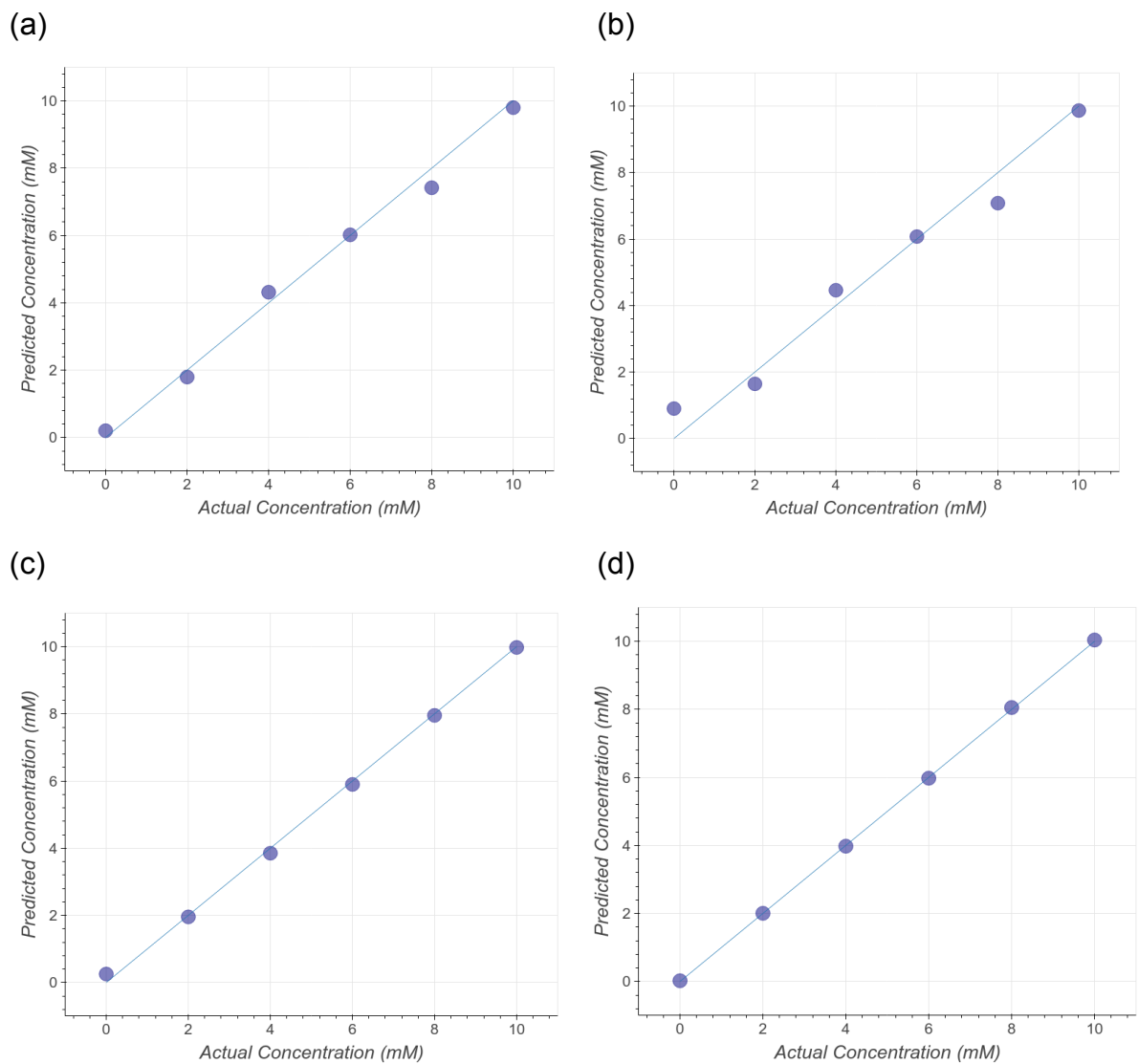

Figure 6: PLS model predicted glucose concentration with respect to actual concentration using the full spectrum of data from each of the four measurement

methods: (a) Photon counter spontaneous ($R^2 = 0.9995$), (b) photon counter stimulated ($R^2 = 0.987$), (c) lock-in spontaneous ($R^2 = 0.9998$), (d) lock-in stimulated ($R^2 = 0.9998$). The greater the strength of the linear regression fit, the lower the mean squared error, and the stronger the predictive power of the PLS model.

Photon Counter

| Range | Spontaneous | Stimulated |
|---|---|---|
| 289-1262 cm$^{-1}$ | 0.0141 mM | 0.694 mM |
| 458-625 cm$^{-1}$ | 0.0933 mM | 0.337 mM |
| 642-1139 cm$^{-1}$ | 0.0467 mM | 0.806 mM |

Lock-in Amplifier

| Range | Spontaneous | Stimulated |
|---|---|---|
| 289-1262 cm$^{-1}$ | $1.70 \times 10^{-2}$ mM | $9.96 \times 10^{-4}$ mM |
| 458-625 cm$^{-1}$ | $7.02 \times 10^{-3}$ mM | $1.84 \times 10^{-3}$ mM |
| 642-1139 cm$^{-1}$ | $1.73 \times 10^{-2}$ mM | $1.04 \times 10^{-3}$ mM |

Table 1: Mean squared error between predicted and actual glucose concentrations for each measurement method over three different spectral ranges. Lower mean squared error values indicate stronger predictive power of the model, so the PLS model based on lock-in stimulated data was consistently the strongest predictive model.

The mean squared error between predicted and actual glucose concentrations for the PLS model based on the full spectral range of lock-in stimulated Raman data is as low as $9.96 \times 10^{-4}$ mM, which is the lowest mean squared error from any of the models based on data from any of the measurements methods. The mean squared error values from the lock-in stimulated Raman data are about an order of magnitude smaller than those from either of the spontaneous Raman data and over 2 orders of magnitude smaller than the mean squared error values from the photon counter stimulated Raman data. When using the lock-in amplifier to reduce error from the Stokes LED, the PLS models based on stimulated Raman data are consistently the strongest glucose concentration predicting models. Interestingly, the lock-in stimulated data is only slightly increased compared to the spontaneous data, but the PLS model based on stimulated data is significantly stronger, likely due in part to the slightly narrower Raman modes of the stimulated spectra. The stimulated model's higher R-squared value of the linear regression between predicted and actual concentration further corroborate this result. Furthermore, when the training, validation, and test sets are switched around; for example, the second set is training, third is validation, and first is test; the results are similar, showing that the lock-in stimulated model is robust and consistent.

**Discussion**

This study demonstrates the viability broadband stimulated Raman with an LED Stokes source and shows the stronger predictive power of a PLS model based on this stimulated Raman data compared to spontaneous Raman when using a lock-in amplifier to reduce noise. With further developments, this technology may be useful for improved cost-effective spectroscopic molecular identification and concentration prediction in a variety of fields. Prediction of glucose concentration has particularly useful biomedical applications in non-invasive blood glucose level monitoring. However, it is important to note that this technology has several limitations, and the enhancement of Raman modes achieved by stimulated Raman gain this way is relatively small compared to traditional SRS, broadband FSRS, and other enhancement technologies, such as SERS. While broadband LED stimulated Raman is intended to be a cost-effective solution and does not necessarily need to match these powerful Raman enhancement technologies, our experiment demonstrates stimulated Raman having only a small enhancement over spontaneous Raman spectroscopy, so further improvements are necessary to be truly useful.

The detection system, specifically the monochromator and photomultiplier tube, could be improved to make taking spectral measurements more practical while removing sources of signal loss. We decided to use the photomultiplier because of its high sensitivity and dynamic range. This was necessary because the unfiltered light from the Stokes LED would saturate a detector with lower dynamic range, and we needed to detect the relatively small stimulated Raman signal on top of the LED signal. Due to using a photomultiplier tube, a monochromator was necessary to isolate a single wavelength, so measuring a full spectrum was accomplished one point at a time. This made taking measurements a time intensive process, which from a practical standpoint, offset the benefit of stimulated Raman gain. Furthermore, our signal experienced significant loss through the monochromator. Replacing the monochromator and photomultiplier with a high dynamic range spectrometer would dramatically speed up the measurement process and potentially increase overall signal to noise ratio.

The LED introduces several challenges to the Raman system. As an incoherent light source, the LED light diffracts quickly, and loses a significant amount of signal when coupled into the multi-furcated fiber. This results in a Stokes source orders of magnitude lower in intensity than the pump at the sample, which leads to low stimulated Raman gain. The LED also adds noise though fluctuations and intensity drift, which may counteract the benefit of creating stimulated Raman. Since the intensity of the Stokes LED is almost 2 orders of magnitude higher than that of the stimulated Raman signal, even the Poisson noise of the LED is significant relative of the stimulated signal. This effect of noise from both the LED and laser was greatly mitigated by using an optical chopper to modulate the pump beam and detecting the stimulated Raman gain signal with a lock-in amplifier synchronized to the modulation of the pump. However, using a lock-in amplifier adds a significant cost to the system, especially if a multi-channel lock-in amplifier would be used to measure a full spectrum all at once. Given cost-effectiveness is one of the main goals of this LED-based stimulated Raman system, the high cost of multi-channel lock-in amplifiers is currently a significant barrier to the practical application of this technology, but the development of cost-effective digital lock-in amplifiers is an active area of research.[33,34] It may also be possible to use the relatively noisy data collected without the lock-in amplifier, using a photon counter in our case or potentially a spectrometer, if the statistical models used were more robust to noise than our partial least squares model. This could be done, in part, by designing a model that penalizes overfitting.

We designed our stimulated Raman system for monitoring glucose concentrations rather than imaging, but broadband LED stimulated Raman also has potential imaging applications. However, aberrations of the LED light would have to be considered, and the imaging resolution would be limited by LED spot size, which is slightly larger than the laser spot size in our setup. Biomedical imaging would require further modifications including lowering the laser power to prevent burning of biological tissue and using NIR wavelengths instead of visible for greater depth of penetration.

**Supplementary Experiment**

Temperature changes in the glucose solution sample could potentially alter the Raman spectra, so we conducted a series of experiments to test whether or not the long exposure of the focused laser and LED beams had any effect on the resulting Raman measurements, either spontaneous or stimulated. Using the 10 mM glucose sample and the same integration time of 10 seconds as in the other experiments, we first continuously measured the spontaneous Raman intensity at a single frequency shift value for 10 minutes, for a total of 120 individual measurements. Then, we measured the stimulated Raman intensity for 10 minutes at the same frequency shift value by subtracting the alternating combined and Stokes measurements as usual. Given each stimulated Raman measurement is the difference of two consecutive measurements, this resulted in 60 stimulated Raman measurements total. This process was performed at 592, 886, 1124, 2101, and 3400 $cm^{-1}$. The Raman modes at 592 and 1124 $cm^{-1}$ are two of the most prominent in the spectra we collected, with that at 592 $cm^{-1}$ being the global maximum and that at 1124 $cm^{-1}$ being the mode which experienced the greatest increase between stimulated Raman and spontaneous Raman. We decided to also take measurements at 886 and 2101 $cm^{-1}$ in order to test if there was a temperature effect on the baseline or off-peak regions, and the highest point of the strong group of water modes, 3400 $cm^{-1}$, was chosen to see if the spectral features related to water would change. Since we were testing the possibility of the laser heating the samples and effecting the Raman measurements, there was a 10 minute gap between rounds of measurements so the sample could cool if needed. Three rounds of measurements were taken at each of the five frequency shift values.

Our experiments showed no significant time-dependent Raman intensity change in any of the tests, suggesting that whatever heating effect the beams may have had on the samples was not significant enough to alter the Raman spectra. This is important because it rules out heating as a potential systematic source of error for our spectral measurements and allows us to make fair comparisons between our spontaneous and stimulated Raman measurements.

**References**

[1] N. C. Dingari, I. Barman, G. P. Singh, J. W. Kang, R. R. Dasari, and M. S. Feld, "Investigation of the specificity of Raman spectroscopy in non-invasive blood glucose measurements," *Anal. Bioanal. Chem.*, vol. 400, no. 9, pp. 2871–2880, Jul. 2011.
[2] P. Nandakumar, A. Kovalev, and A. Volkmer, "Vibrational imaging based on stimulated Raman scattering microscopy," *New J. Phys.*, vol. 11, no. 3, p. 33026, 2009.
[3] J.-X. Cheng and X. S. Xie, "Vibrational spectroscopic imaging of living systems: An emerging platform for biology and medicine," *Science (80-. ).*, vol. 350, no. 6264, pp. aaa8870-aaa8870, 2015.
[4] C.-S. Liao and J.-X. Cheng, "In Situ and In Vivo Molecular Analysis by Coherent Raman Scattering Microscopy," *Annu. Rev. Anal. Chem.*, vol. 9, no. 1, pp. 69–93, 2016.
[5] E. R. Andresen, P. Berto, and H. Rigneault, "Stimulated Raman scattering microscopy by spectral focusing and fiber-generated soliton as Stokes pulse," *Opt. Lett.*, vol. 36, no. 13, p. 2387, 2011.
[6] D. Zhang, M. N. Slipchenko, and J. X. Cheng, "Highly sensitive vibrational imaging by femtosecond pulse stimulated raman loss," *J. Phys. Chem. Lett.*, vol. 2, no. 11, pp. 1248–1253, 2011.
[7] B. G. Saar, C. W. Freudiger, J. Reichman, C. M. Stanley, G. R. Holtom, and X. S. Xie, "Video-Rate Molecular Imaging in Vivo with Stimulated Raman Scattering.," *Sci. (Washington, DC, United States)*, vol. 330, no. 6009, pp. 1368–1370, 2010.
[8] C. W. Freudiger, W. Min, G. R. Holtom, B. Xu, M. Dantus, and X. S. Xie, "Highly specific label-free molecular imaging with spectrally tailored excitation-stimulated Raman scattering (STE-SRS) microscopy," *Nat. Photonics*, vol. 5, no. 2, pp. 103–109, 2011.
[9] M. Ji *et al.*, "Detection of human brain tumor infiltration with quantitative stimulated Raman scattering microscopy," *Sci. Transl. Med.*, vol. 7, no. 309, p. 309ra163-309ra163, 2015.
[10] C. S. Liao *et al.*, "In Vivo and in Situ Spectroscopic Imaging by a Handheld Stimulated Raman Scattering Microscope," *ACS Photonics*, vol. 5, no. 3, pp. 947–954, 2018.
[11] T. W. Kee and M. T. Cicerone, "Simple approach to one-laser, broadband coherent anti-Stokes Raman scattering microscopy," *Opt. Lett.*, vol. 29, no. 23, pp. 2701–2703, 2004.
[12] H. T. Beier, G. D. Noojin, and B. A. Rockwell, "Stimulated Raman scattering using a single femtosecond oscillator with flexibility for imaging and spectral applications," *Opt. Express*, vol. 19, no. 20, p. 18885, 2011.
[13] K. Yen, T. T. Le, A. Bansal, S. D. Narasimhan, J. X. Cheng, and H. A. Tissenbaum, "A comparative study of fat storage quantitation in nematode Caenorhabditis elegans using label and label-free methods," *PLoS One*, vol. 5, no. 9, pp. 1–10, 2010.
[14] Y. Ozeki, F. Dake, S. Kajiyama, K. Fukui, and K. Itoh, "Analysis and experimental assessment of the sensitivity of stimulated Raman scattering microscopy," *Opt. Express*, vol. 17, no. 5, p. 3651, 2009.
[15] W. Min, C. W. Freudiger, S. Lu, and X. S. Xie, "Coherent Nonlinear Optical Imaging: Beyond Fluorescence Microscopy," *Annu. Rev. Phys. Chem.*, vol. 62, no. 1, pp. 507–530, 2011.
[16] J. Shao *et al.*, "In Vivo Blood Glucose Quantification Using Raman Spectroscopy," *PLoS One*, 2012.
[17] P. N. Malevich *et al.*, "Ultrafast-laser-induced backward stimulated Raman scattering for tracing atmospheric gases," *Opt. Express*, vol. 20, no. 17, p. 18784, 2012.
[18] T. Steinle *et al.*, "Synchronization-free all-solid-state laser system for stimulated Raman scattering microscopy," *Light Sci. Appl.*, vol. 5, no. 10, pp. 1–6, 2016.
[19] Y. Ozeki, W. Umemura, K. Sumimura, N. Nishizawa, K. Fukui, and K. Itoh, "Stimulated Raman hyperspectral imaging based on spectral filtering of broadband fiber laser pulses," *Opt. Lett.*, vol. 37, no. 3, p. 431, 2012.

[20] D. Fu et al., "Quantitative Chemical Imaging with Multiplex Stimulated Raman Scattering Microscopy," *J. Am. Chem. Soc.*, vol. 134, no. 8, pp. 3623–3626, 2012.

[21] Z. Meng, G. I. Petrov, and V. V. Yakovlev, "Pure electrical, highly-efficient and sidelobe free coherent Raman spectroscopy using acousto-optics tunable filter (AOTF)," *Sci. Rep.*, vol. 6, no. February, pp. 1–7, 2016.

[22] D. W. McCamant, P. Kukura, S. Yoon, and R. A. Mathies, "Femtosecond broadband stimulated Raman spectroscopy: Apparatus and methods," *Rev. Sci. Instrum.*, vol. 75, no. 11, pp. 4971–4980, 2004.

[23] P. Kukura, D. W. McCamant, and R. A. Mathies, "Femtosecond Stimulated Raman Spectroscopy," *Annu. Rev. Phys. Chem.*, vol. 58, no. 1, pp. 461–488, 2007.

[24] E. Ploetz, S. Laimgruber, S. Berner, W. Zinth, and P. Gilch, "Femtosecond stimulated Raman microscopy," *Appl. Phys. B Lasers Opt.*, vol. 87, no. 3, pp. 389–393, 2007.

[25] K. Seto, Y. Okuda, E. Tokunaga, and T. Kobayashi, "Development of a multiplex stimulated Raman microscope for spectral imaging through multi-channel lock-in detection," *Rev. Sci. Instrum.*, vol. 84, no. 8, 2013.

[26] S. K. Vashist, "Non-invasive glucose monitoring technology in diabetes management: A review," *Anal. Chim. Acta*, vol. 750, pp. 16–27, 2012.

[27] M. S. Wróbel, "Non-invasive blood glucose monitoring with Raman spectroscopy: Prospects for device miniaturization," *IOP Conf. Ser. Mater. Sci. Eng.*, vol. 104, no. 1, 2016.

[28] D. M. Haaland and E. V. Thomas, "Partial Least-Squares Methods for Spectral Analyses. 1. Relation to Other Quantitative Calibration Methods and the Extraction of Qualitative Information," *Anal. Chem.*, vol. 60, no. 11, pp. 1193–1202, 1988.

[29] K. P. J. Williams and N. J. Everall, "Use of micro Raman spectroscopy for the quantitative determination of polyethylene density using partial least-squares calibration," *J. Raman Spectrosc.*, vol. 26, no. 6, pp. 426–433, 1995.

[30] M. Mathlouthi and D. Vinh Luu, "Laser-raman spectra of d-glucose and sucrose in aqueous solution," Carbohydr. Res., vol. 81, no. 2, pp. 203–212, 1980.

[31] P. D. Vasko, J. Blackwell, and J. L. Koenig, "Infrared and raman spectroscopy of carbohydrates.," Carbohydr. Res., vol. 23, no. 3, pp. 407–416, 1972.

[32] Scikit-learn: Machine Learning in Python, Pedregosa et al., JMLR 12, pp. 2825-2830, 2011.

[33] A. Kar and M. Chandra, "A Low-Cost , Portable Alternative for a Digital Lock- In Amplifier Using TMS320C5535 DSP," 2014.

[34] J. Lu, D. Pan, and L. Qiao, "The principle of a virtual multi-channel lock-in amplifier and its application to magnetoelectric measurement system," Meas. Sci. Technol., vol. 045702, no. iii, p. 11, 2007.